\documentclass[fleqn,usenatbib]{mnras}

\usepackage{newtxtext,newtxmath}

\usepackage[T1]{fontenc}

\DeclareRobustCommand{\VAN}[3]{#2}
\let\VANthebibliography\thebibliography
\def\thebibliography{\DeclareRobustCommand{\VAN}[3]{##3}\VANthebibliography}

\usepackage{graphicx}	
\usepackage{amsmath}	
\usepackage{enumitem}
\usepackage{siunitx}
\usepackage{booktabs}
\usepackage{dsfont}
\usepackage{orcidlink}
\usepackage{balance}

\DeclareSIUnit\parsec{pc}
\DeclareSIUnit\megaparsec{Mpc}
\DeclareSIUnit\h{h}

\title{Systematic assessment of the Hubble tension via Bayesian jackknife testing}

\author[T. Hughes et al.]{
Thomas Hughes \orcidlink{https://orcid.org/0009-0002-7123-0156},$^{1}$ 
Michael J. Wilensky \orcidlink{https://orcid.org/0000-0001-7716-9312},$^{2,1, \dagger, \parallel}$ 
Philip Bull \orcidlink{https://orcid.org/0000-0001-5668-3101}$^{1,3}$\thanks{E-mail: phil.bull@manchester.ac.uk}
\\
$^{1}$Jodrell Bank Centre for Astrophysics, University of Manchester, Manchester M13 9PL, UK\\
$^{2}$Department of Physics and Trottier Space Institute, McGill University, 3600 University Street, Montreal, QC H3A 2T8, Canada\\
$^{3}$Department of Physics and Astronomy, University of Western Cape, Cape Town 7535, South Africa\\
$^{\dagger}$CITA National Fellow\\
$^{\parallel}$TSI Postdoctoral Fellow\\
}

\date{Accepted XXX. Received YYY; in original form ZZZ}

\pubyear{\the\year{}}

\begin{document}
\label{firstpage}
\pagerange{\pageref{firstpage}--\pageref{lastpage}}
\maketitle

\begin{abstract}
Statistically-significant differences in the value of the Hubble parameter are found depending on the measurement method that is used, a result known as the Hubble tension. A variety of ways of comparing, grouping, and excluding measurements have been used to try to explain this, either in terms of physical effects or systematic errors. We present a systematic `Bayesian jackknife' analysis of 16 independent measurements of the Hubble parameter in an attempt to identify whether the measurements fall into meaningful clusters that would help explain the origin of the tension. 
After evaluating evidence ratios for the commonly-used split into early- vs late-time measurements, we then study a range of simplified alternative physical scenarios that reflect different physical origins of an apparent bias or shift in the value of $H_0$, assigning phenomenological population parameters to each subset. These include scenarios where specific subsets are biased (e.g. due to unrecognised experimental systematics in the local distance ladder or cosmic microwave background measurements), as well as more cosmologically-motivated cases involving modifications to the expansion history. Many of these scenarios have similar marginal likelihood, but the model where no measurements are biased is strongly disfavoured. Finally, we marginalise over all these scenarios to estimate the `model agnostic' posterior distribution of $H_0$. The resulting distribution is mildly multi-modal, but modestly favours values near $H_0=68$ \si{\km\per\second\per\mega\parsec}, with a 95\% credible region of $66.7 < H_0 < 72.7$ \si{\km\per\second\per\mega\parsec}.

\end{abstract}




\section{Introduction}

Recent observational advances have led to increasingly precise cosmological parameter measurements. Rather than resolving open questions, such as the nature of dark matter or dark energy, this progress has resulted in discrepancies and inconsistencies however, most notably a significant tension between different measurements of the Hubble parameter $H_0$. Local (late Universe) determinations using Cepheid-calibrated supernovae yield values of $H_0 =73.04\pm1.04$ \si{\km\per\second\per\mega\parsec} \citep{Riess_2022Hubble} or similar, while early Universe estimates inferred from cosmic microwave background (CMB) anisotropies under the $\Lambda$CDM model are consistent with values closer to $H_0 = 67.27\pm0.6$ \si{\km\per\second\per\mega\parsec} \citep{Planck2018Parameters}. The $\Lambda$CDM model, which assumes a flat universe dominated by cold dark matter and a cosmological constant, has long served as the standard model for interpreting astrophysical data, making this disagreement particularly consequential. This discrepancy, dubbed the ``Hubble tension,'' has now surpassed the $5\sigma$ threshold, making it one of the most statistically significant challenges to the standard cosmological model. It remains unresolved despite substantial advances in measurement techniques and calibration precision \citep{DES2023LCDM, freedman2024statusreportchicagocarnegiehubble, lee2024chicagocarnegiehubbleprogramjwst, Adame2024}.

The tension raises a crucial question: is the mismatch indicative of new physics, or does it reflect underestimated systematic errors in some of the datasets? Some have argued that for local measurements, unaccounted for systematics such as peculiar velocities \citep{Mukherjee2021, Dalang_2023}, calibration errors \citep{Scolnic2024}, or heliocentric redshift uncertainties \citep{Rameez_2021}, could be inflating estimates of $H_0$. Intrinsic scatter in Type Ia supernova luminosities further complicate the picture; simplified models that attribute this scatter to a single hidden variable may neglect subtler dependencies on dust extinction and host galaxy properties \citep{Wojtak2022, Meldorf2023, Lee2022}, thereby introducing hidden biases in inferred distances. An alternative perspective focuses on issues within the local distance ladder (LDL) itself: while the third rung (Type Ia supernovae) has been highlighted as potentially problematic \citep{Huang2024}, lingering systematics in the earlier calibration steps, parallax and Cepheid distance, may also contribute \citep{LIU2024138717, Scolnic2024}. Others have suggested that assumptions embedded in early Universe physics, including recombination modelling \citep{Davies2019Hubble, Lynch2024, banik2024constraintshubblematterdensity}, incomplete sky coverage \citep{Macpherson_2024}, or exotic components such as primordial magnetic fields \citep{jedamzik2023primordialmagneticfieldshubble}, might bias CMB-based inferences. Several theoretical models have also emerged to resolve the tension through new physics. Early Dark Energy (EDE) models, which propose an additional energy component that is dynamically important before recombination, have emerged as one of the most promising avenues, as they allow higher inferred values of $H_0$ from early Universe probes without significantly disturbing other cosmological parameters \citep{Poulin_2019, DiValentino2021, Lynch2024}. Additional theories involve interactions between dark matter and dark energy, modifications to neutrino properties such as mass hierarchies or sterile species, or the inclusion of novel forms of exotic energy \citep{Weikang2017Discordance, Davies2019Hubble, yeung2024resolvingh0s8tensions, Aboubrahim_2024, Kou2024, Clifton_2024}. Still, even among these, many models provide only partial relief from the tension, failing to provide a complete resolution.

Framing the problem strictly as ``early versus late'' may be an oversimplification. Several methods, including time-delay lensing \citep{Li2024}, baryon acoustic oscillations (BAOs) \citep{DES2023LCDM}, and gravitational wave sirens \citep{Sneppen_2023}, depend indirectly on the sound horizon and thus do not fit neatly into either category. As a result, alternative framings have been proposed, such as comparisons based on methodology rather than redshift alone. In this context, a useful distinction emerges within local measurements, which can be split into two categories: LDL approaches and one-step methods. LDL techniques estimate $H_0$ by calibrating distances using a chain of calibration `rungs', most commonly using geometric parallax to Cepheids and then SNIa \citep{Riess_2021, Riess_2022Hubble}, each introducing potential systematics. In contrast, one-step methods aim to bypass intermediate calibrations entirely and infer $H_0$  directly from cosmological phenomena. Some examples include strong lensing time delays \citep{Li2024}, standard sirens \citep{gayathri2020hubbleconstantmeasurementgw190521} and the Sunyaev-Zel'dovich (SZ) effect \citep{Wan2021}.

The complexity of the Hubble tension has led to a number of statistical approaches being proposed. Initial efforts using tension metrics \citep{Lemos2020Assessing}, while insightful, showed sensitivity to assumptions about model priors and data independence. These insights build on early foundational work by \cite{Press1996Bayesian}, who advocated for hierarchical Bayesian approaches in cosmology, emphasising its capacity to propagate uncertainties, incorporate prior information and perform robust model comparisons. More recently, the formalisation of these methods in the context of the Hubble tension has been advanced by \citet{Feeney2018}, who found that early Universe and late Universe measurements yield significantly different $H_0$ values. The tension is ostensibly reduced with this approach to $3\sigma$, but still remains strong.

In light of this, new Bayesian statistical methods have emerged. \cite{Wilensky2022Bayesian} introduced \texttt{CHIBORG},\footnote{\url{https://github.com/mwilensky768/chiborg}} a `Bayesian jackknife' method, which uses Bayesian inference and combinatorics to identify subsets of data that may be responsible for introducing a bias into a measurement. An analysis of $21$cm power spectrum measurements from the Hydrogen Epoch of Reionization Array (HERA) showed how Bayesian tests can improve reliability by validating which data subsets are most consistent with one another \citep{HERA2023}. \texttt{CHIBORG} uses the \textit{Bayesian evidence} \citep[marginal likelihood;][]{Kass01061995} to compare various hypotheses regarding potential biases in the data. 

In this paper, we extend the work of \cite{Wilensky2022Bayesian} by applying \texttt{CHIBORG} to independent $H_0$ measurements. This allows us to perform systematic model comparisons to assess whether certain data subsets are the main drivers of the tension. In effect, this constitutes a jackknife procedure: the data are split into subsets to see if there is a certain pattern, and here the subsets (and their statistical properties) are defined by the hypotheses. Through this, we aim to isolate potentially problematic measurements that may be introducing a systematic bias.

We first examine early versus late Universe measurements of $H_0$, as well as LDL versus one-step methods, using \texttt{PolyChord} to perform the model comparison. \texttt{PolyChord} is a nested sampler that uses slice sampling and dimensional reduction techniques to calculate the Bayesian evidence, but also generates posterior samples as a natural by-product \citep{Handley2015b, Handley2015a}. We then explore a suite of alternative phenomenological models with realistic bias structures to see whether the data can more specifically identify the source of the tension. Finally, we marginalise over these models to produce a refined `model agnostic' estimate of $H_0$, and investigate how excluding suspect data points impacts the inferred value.

In Section \ref{sec: maths}, we present the mathematical formalism for evidence calculations, which forms the basis for the Bayesian jackknife analysis. Section \ref{sec: methods} describes the methodology for implementing the jackknife tests using \texttt{CHIBORG}, as well as the model comparison employed in \texttt{PolyChord}. In Section \ref{sec: Hypotheses}, we detail the set of hypotheses tested in our analysis. Section \ref{sec: results} reports the results of both the jackknife and model comparison tests, along with a discussion of their implications. Finally, in Section \ref{sec: conclusion}, we summarise our findings and outline the conclusions drawn from this work.

\section{Mathematical Formalism}
\label{sec: maths}

\begin{figure*}
    \centering
    \includegraphics[width=1.65\columnwidth]{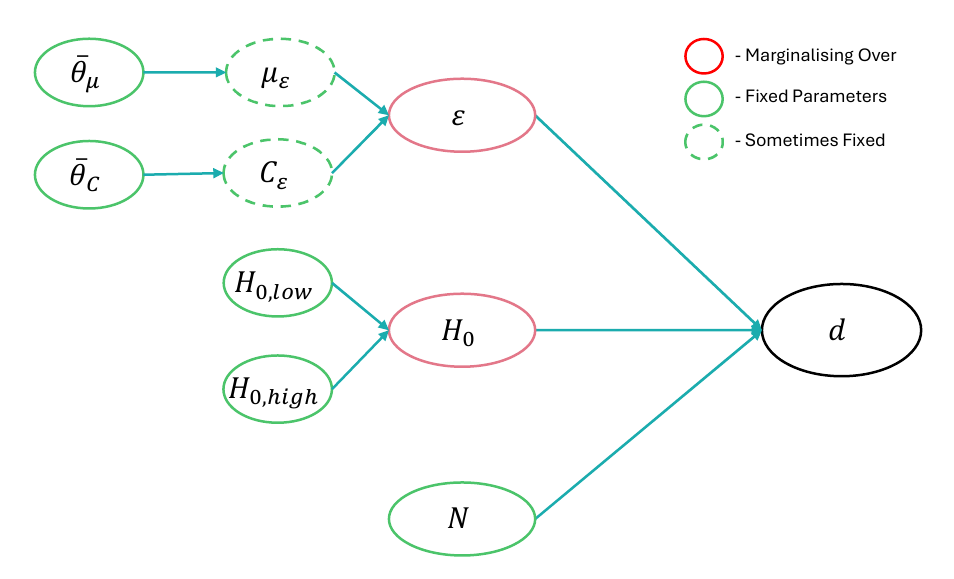}
    \caption{Parameter structure used in the initial model comparison, illustrating the relationships between hyperparameters, model parameters, and the data vector \(\mathbf{d}\). Parameters shown in green are fixed, while those in red are marginalised over in the analysis. Parameters shown in dashed green lines are sometimes fixed, depending on the test. In later phenomenological scenarios, \(\bar{\theta}_\mu\) and \(\bar{\theta}_C\) are removed, with \(\mu_\varepsilon\) and \(C_\varepsilon\) fixed throughout the analysis.}
    \label{fig:chiborg dag}
\end{figure*}

In this section, we outline the principles of Bayesian inference and model comparison used throughout this work. Our goal is to assess the relative plausibility of competing hypotheses about which $H_0$ measurements may be systematically biased. Bayesian model comparison allows us to evaluate competing hypotheses by defining a model corresponding to each hypothesis and comparing how well each model explains the data while accounting for model complexity. Given a model $\mathcal{M}_j$ with parameters $\theta$, Bayes’ theorem provides the posterior distribution over those parameters conditional on data, $\mathbf{d}$, as
\begin{equation}
P(\theta | \mathbf{d}, \mathcal{M}_j) = \frac{P(\mathbf{d} | \theta, \mathcal{M}_j) P(\theta | \mathcal{M}_j)}{P(\mathbf{d} | \mathcal{M}_j)} 
\end{equation}
where $P(\mathbf{d} | \theta, \mathcal{M}_j)$ is the likelihood, $P(\theta | \mathcal{M}_j)$ is the prior probability distribution for the parameter values within model $\mathcal{M}_j$, and the denominator, known as the Bayesian evidence (or marginal likelihood) $\mathcal{Z}$, normalises the posterior. The evidence is defined by
\begin{equation}
\mathcal{Z} = P(\mathbf{d} | \mathcal{M}_j) = \int P(\mathbf{d} | \theta, \mathcal{M}_j) .P(\theta | \mathcal{M}_j)  d\theta 
\label{eq:evidence}
\end{equation}
This captures both the fit of the model to the data and its ability to generalise over the prior space, making it central to model comparison. The posterior probability of a given model $\mathcal{M}_j$ is determined via Bayes' theorem,
\begin{equation}
\label{eq:bayes_theorem}
P(\mathcal{M}_j | \mathbf{d}) = \frac{P(\mathbf{d} | \mathcal{M}_j) \, P(\mathcal{M}_j)}{P(\mathbf{d})}.
\end{equation}
Here, $P(\mathbf{d} | \mathcal{M}_j)$ is the marginal likelihood for model $\mathcal{M}_j$, $P(\mathcal{M}_j)$ is the prior for model $\mathcal{M}_j$ and the denominator $P(\mathbf{d})$ ensures proper normalisation over the model space and is equal to the sum of the marginal likelihoods weighted by their respective priors. This formulation allows for comparison between multiple hypotheses by quantifying their relative posterior probabilities.

To evaluate the evidence \( \mathcal{Z} \) for each model, we require a joint probability density of (noisy) $H_0$ measurements, represented by the vector $\mathbf{d}$, given a fixed value of $H_0$. This probability density defines a likelihood function for $H_0$. Assuming Gaussian distributions for both the bias and measurement noise, this likelihood can be derived analytically. Following \citet{Wilensky2022Bayesian}, we model the data as
\begin{equation}
\label{eq:data_model}
\mathbf{d} \mid H_0, \boldsymbol{\varepsilon}, \mathbf{n} = H_0 \mathds{1} + \boldsymbol{\varepsilon} + \mathbf{n} 
\end{equation}
where `=' represents a deterministic relationship, so once $H_0$, the systematic biases $\boldsymbol{\varepsilon}$, and the measurement noise $\mathbf{n}$ are specified, the observed data vector $\mathbf{d}$ is fully determined. $\mathbf{n}$ is a random error in each measurement consistent with its quoted statistical uncertainties, while $\boldsymbol{\varepsilon}$ is an additional systematic error that is not captured by these. $\mathds{1}$ is a vector where every component is equal to 1.

To perform model comparison, we need to establish statistical models (i.e. priors) for $H_0$, $\boldsymbol{\varepsilon}$, and $\mathbf{n}$. To facilitate analytic marginalization, we suppose all quantities except for $H_0$ are Gaussian \textit{a priori}. Some of the statistical uncertainties from various measurements are non-Gaussian, and in particular some of them have skewed distributions indicated by uneven upper and lower error bars in Table \ref{tab:16 Hubble measurements}. To treat these skewed uncertainties as Gaussian, we take the average of their upper and lower reported error bar, and square it to obtain the noise variance. We denote this variance as a diagonal matrix, $\mathbf{N}$.

For each model, we define a bias mean vector, $\boldsymbol{\mu}_\varepsilon$, and covariance matrix, $\mathbf{C}_\varepsilon$. These serve as population parameters for different subsets of the data. The structure of these parameters determines the nature of the bias. For example, a diagonal $\mathbf{C}_\varepsilon$ would suggest that there is excess variance in some measurements beyond the quoted statistical uncertainties. This would appear to cause independent fluctuations from measurement to measurement, which might be expected from an experimental systematic that varies between experiments.

Supposing nonzero off-diagonal elements implies that the effect of the systematic has a certain level of consistency from experiment to experiment. As another example, if some subset of data are considered (exactly) unbiased within some model, then corresponding blocks of $\boldsymbol{\mu}_\varepsilon$ and $\mathbf{C}_\varepsilon$ are set to 0, and so on. In any case, the Gaussian specification of $\boldsymbol{\varepsilon}$ means we can marginalise over it (and the noise term) analytically to obtain
\begin{equation}
    \mathbf{d} \mid H_0, \boldsymbol{\mu}_{\varepsilon}, \mathbf{C}_{\varepsilon} \sim \mathcal{MN}(H_0 \mathds{1} + \boldsymbol{\mu}_{\varepsilon}, \mathbf{C}_{\varepsilon} + \mathbf{N}),
    \label{eq:fix_like}
\end{equation}
where $\mathbf{x} \sim \mathcal{MN}\left(\boldsymbol{\mu}, \boldsymbol{\Sigma}\right)$ denotes that the vector $\mathbf{x}$ has multivariate normal distribution with mean $\boldsymbol{\mu}$ and covariance $\boldsymbol{\Sigma}$. 

In our first stage of model comparison (early vs. late, \S\ref{sec: early vs late}), we consider the overall amplitudes of $\boldsymbol{\mu}_\varepsilon$ and $\mathbf{C}_\varepsilon$ as unknown, and marginalise over this uncertainty. This demands priors for these amplitudes, with hyperparameters denoted $\overline{\theta}_\mu$ and $\overline{\theta}_C$, which are discussed in more detail in \S\ref{sec: early vs late}. The full marginalization is analytically intractable, and so we use \texttt{PolyChord} to calculate the Bayesian evidence (Eq.~\ref{eq:evidence}). In our second round of model comparison (physical hypotheses, \S\ref{sec: multiple hypotheses}), we fix $\boldsymbol{\mu}_\varepsilon$ and $\mathbf{C}_\varepsilon$ to particular values according to the given hypothesis, allowing the use of Eq.~\ref{eq:fix_like} to define a likelihood for $H_0$ directly. We discuss how these values are fixed in \S\ref{sec: multiple hypotheses}. While analytic marginalization over a uniform $H_0$ prior is tractable with this likelihood, we use \texttt{PolyChord} since the output posterior samples make it easy to establish a model-marginalised posterior probability distribution for the true value of $H_0$. We also note that a Gaussian prior on $H_0$ with fixed $\boldsymbol{\mu}_\varepsilon$ and $\mathbf{C}_\varepsilon$ allows for a fully analytic marginalization, presented in Appendix \ref{Appendix A}, which recovers the result from \citet{Wilensky2022Bayesian} but with a much simpler derivation. 

We summarise the parametric dependencies in this analysis via a graphical representation in Figure \ref{fig:chiborg dag}. Parameters shown in red are marginalised over in the inference, while those in solid green are always held fixed. As mentioned previously, some jackknife tests involve marginalizing over $\boldsymbol{\mu}_\varepsilon$ and $\mathbf{C}_\varepsilon$, while some involve fixing them to certain values in a model-dependent way. We denote the variables affected by this choice with dashed green outlines, and defer discussion of particular choices to \S\ref{sec: Hypotheses}. 

\section{Methods}
\label{sec: methods}

In this section, we summarise the methodological steps taken to prepare for the model comparison analysis. This includes describing the \texttt{CHIBORG} code, assembling a set of independent $H_0$ measurements, and describing the use of \texttt{PolyChord} to allow for inference/marginalisation of free parameters.

\begin{table*}
    \centering
    \setlength{\tabcolsep}{2pt}
    \begin{tabular}{c l c c c c c c l}
        \toprule 
        & \textbf{Measurement} & \textbf{Measured $H_0$} & \textbf{Lower Error} & \textbf{Upper Error} & \textbf{Category} & \textbf{Early/Late} & \textbf{Redshift} & \textbf{Reference} \\
        & & (\si{\km\per\second\per\mega\parsec}) & (\si{\km\per\second\per\mega\parsec}) & (\si{\km\per\second\per\mega\parsec}) & & \textbf{Universe} & ($z$)\\ 
        \midrule \midrule
        1 & CMB with Planck & 67.3 & 0.6 & 0.6 & CMB & Early & Distant & \cite{Planck2018Parameters}\\ 
        2 & CMB without Planck & 67.9 & 1.5 & 1.5 & CMB & Early & Distant & \cite{Aiola_2020}\\ 
        3 & BAO & 68.5 & 2.2 & 2.2 & BAO & Early & Distant & \cite{d_Amico_2020}\\ 
        4 & Cepheids-SNIa & 73.2 & 1.3 & 1.3 & LDL & Late & Local & \cite{Riess_2021}\\ 
        5 & TRGB-SNIa & 72.1 & 2.0 & 2.0 & LDL & Late & Local & \cite{Soltis_2021}\\ 
        6 & JAGB-SNIa & 68.0 & 2.7 & 2.7 & LDL & Late & Local & \cite{lee2024chicagocarnegiehubbleprogramjwst}\\ 
        7 & Tully-Fisher Relation & 76.0 & 2.6 & 2.6 & LDL & Late & Local & \cite{Kourkchi_2020}\\ 
        8 & Surface Brightness Fluctuations  & 73.3 & 2.5 & 2.5 & LDL & Late & Local & \cite{Blakeslee_2021}\\ 
        9 & SNII & 75.8 & 4.9 & 5.2 & LDL & Late & Local & \cite{de_Jaeger_2020}\\ 
        10 & HII galaxies & 71.0 & 3.5 & 3.5 & LDL & Late & Local & \cite{Fern_ndez_Arenas_2017}\\ 
        11 & Gravitational Waves & 73.4 & 10.7 & 6.9 & One-step & Late & Distant & \cite{gayathri2020hubbleconstantmeasurementgw190521}\\ 
        12 & Fast Radio Bursts (FRBs) & 64.7 & 4.7 & 5.6 & One-step & Late & Distant & \cite{Wu_2022}\\ 
        13 & EPM + Kilonovae & 68.0 & 3.6 & 3.6 & One-step & Late & Local & \cite{Sneppen_2023}\\ 
        14 & Gamma ray attenuation & 62.4 & 3.9 & 4.1 & One-step & Late & Distant & \cite{domínguez2023newderivationhubbleconstant}\\ 
        15 & Time Delay Cosmography & 66.0 & 4.0 & 4.0 & One-step & Late & Distant & \cite{Li2024}\\ 
        16 & Masers & 73.9 & 3.0 & 3.0 & One-step & Late & Local & \cite{Pesce_2020}\\ 
        \bottomrule
    \end{tabular}
    \caption{The 16 independent $H_0$ measurements used in this study. Here, `Local' redshift corresponds to $z\leq0.2$ and `Distant' redshift to $z>0.2$.}
    \label{tab:16 Hubble measurements}
\end{table*}

\begin{figure}
    \centering
    \includegraphics[width=1\columnwidth]{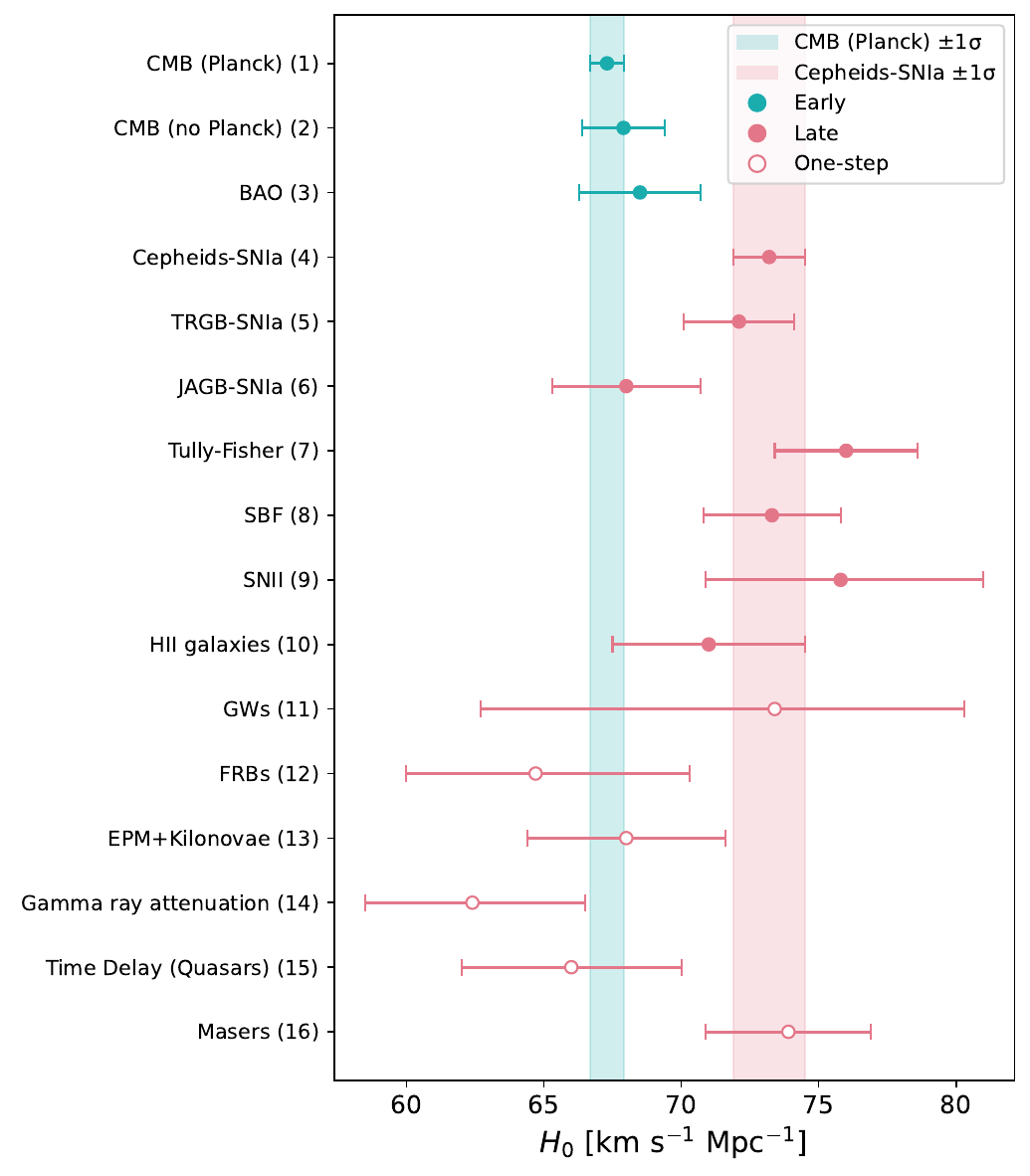}
    \caption{The 16 independent $H_0$ measurements used in this analysis, corresponding to the values listed in Table \ref{tab:16 Hubble measurements}. Early Universe measurements (CMB and BAO) are shown in blue, late Universe distance ladder measurements in red, and one-step methods with open red markers. The shaded bands indicate the $\pm 1\sigma$ regions for the Planck CMB (blue) and Cepheids-SN Ia (red) measurements, which are commonly used as reference values in the Hubble tension discussion.}
    \label{fig:measurements}
\end{figure}

\subsection{\texttt{CHIBORG}}

Throughout this work, we use \texttt{CHIBORG}, an open-source Python package, to perform Bayesian jackknife tests aimed at detecting potential biases within datasets. These tests, introduced in \cite{Wilensky2022Bayesian}, are particularly effective in diagnosing statistical tension between multiple measurements of the same physical quantity. \texttt{CHIBORG} applies Bayesian inference to compute the evidence for competing hypotheses about where bias may exist in the data, allowing for rigorous model comparison. \texttt{CHIBORG} offers two approaches to calculate the Bayesian evidence: a fully analytic method assuming Gaussian priors (shown in Appendix \ref{Appendix A}) and a numerical option that marginalises over a non-Gaussian prior on the `signal', the set of true underlying model parameters, such as $H_0$, using Gaussian quadrature. Conceptually, this serves a similar role to a $\chi^2$ test, but with a key distinction: rather than testing for overall goodness of fit, \texttt{CHIBORG} systematically evaluates many different combinations of data points or groups, identifying which are most likely to be responsible for the statistical tension. 

In practice, our analysis proceeds through a few key steps:
\begin{enumerate}[label=\arabic*., leftmargin=*, labelwidth=0pt]
    \item \textbf{Define models:} Specify hypotheses that include or exclude potential biases for individual measurements or groups.
    \item \textbf{Generate subsets:} Create resampled subsets by systematically leaving out one or more data points.
    \item \textbf{Compute Bayesian evidence:} Evaluate the evidence for each model and subset using either an analytic or numerical approach.
    \item \textbf{Construct posterior probabilities:} Combine evidences to quantify the relative plausibility of different bias scenarios.
\end{enumerate}

Note that step 2 in the above is often referred to as ``jackknife resampling,'' whereas in this work when we say ``jackknife test'' we primarily mean steps 1, 3, and 4, i.e. ``separating the considered data into subsets, proposing different statistical structure for each subset, and evaluating the evidence for or against these different statistical structures.'' However, in preliminary analyses we observed sensitivity to jackknife test results depending on which data were considered, and so we include a resampling analysis in this work to help probe the robustness of our inferences. This is to say, for many resampled subsets of the data, we further break each subset into model-defined subsets and apply the model comparison analysis (i.e. the jackknife test). By examining the jackknife test consistency across many resampled subsets of the data, the method can pinpoint measurements or groups that are particularly critical for driving the inference. This is specific to jackknife tests using the physical hypotheses, and we discuss it in more detail in \S\ref{sec: multiple hyp results}.

\subsection{Data Curation}

We compiled 16 measurements of $H_0$, prioritising independence between datasets. While some $H_0$ measurements in the literature are known to be correlated, we avoid including such data in this analysis as far as possible. Instead, we restrict ourselves to a set of independent measurements, each treated as a Gaussian random variable characterised by its reported central value and quoted uncertainty. The full list of measurements used is shown in Table \ref{tab:16 Hubble measurements} and Fig.~\ref{fig:measurements}.

\subsection{\texttt{PolyChord}}

The original implementation of \texttt{CHIBORG} required that both the bias mean and bias variance be fixed to arbitrary values, which imposed potentially strong assumptions on the analysis. These fixed parameters could influence the inference results by changing the shape of the posterior and reducing the model's ability to reflect realistic systematic uncertainties. To address this limitation and allow for a more data-driven approach, we replaced the fixed-parameter model with a hierarchical model that allows the bias parameters to themselves be considered uncertain. This approach required the use of \texttt{PolyChord}, a nested sampling algorithm well suited for high-dimensional parameter spaces and Bayesian model comparison.

\texttt{PolyChord} has two tunable parameters that affect the output quality: the number of live points, $N_\mathrm{live}$ and the number of slice sampling steps before generating a new live point, $N_\mathrm{repeats}$. We use $N_\mathrm{live}=300$ and $N_\mathrm{repearts} = 5N_\mathrm{dim}$, where $N_\mathrm{dim}$ is the dimensionality of the model in consideration. See \citet{Handley2015a} for a more detailed description of the effects of these parameter choices. In addition to evidence values, \texttt{PolyChord} produces samples from the full joint posterior distribution of all model parameters. This output allows for detailed investigation of how parameters interact within each scenario and we gain insight into the structure and features of the probability distributions, such as multimodality, or parameter degeneracies. Since these outputs include the Bayesian evidence for each model, we can easily marginalise over models to produce a combined $H_0$ posterior from the outputs.

\section{Hypotheses}
\label{sec: Hypotheses}

In this section, we outline a set of hypotheses that could explain the tension in $H_0$ measurements, which are then compared in our analysis. These include comparisons between early and late Universe measurements and a range of alternative phenomenological models. The hypotheses are intended to be simplified `generic' versions of these models that can be implemented using our simple bias formalism, in order to avoid the complexity of including many different model-dependent implementations into the inference. As such, they cannot be considered exhaustive or complete representations of the behaviours of these different model classes. This should be noted as an important caveat to the analysis that follows. However, they are sufficient to illustrate how different types of model should affect the expected $H_0$ values from different observations, essentially by `biasing' them away from the basic $\Lambda$CDM behaviour.

We allow biases to be drawn from a distribution with mean $\mu_\varepsilon$ and variance $\sigma_\varepsilon^2$. For example, a mean of +4 and a standard deviation of 1 would imply that a given data point is likely to be biased by +4 $\pm$ 1 \si{\km\per\second\per\mega\parsec}. Larger or smaller biases than this are allowed, but with reduced probability, following the shape of a Gaussian distribution with these parameters. We allow the bias parameters, the mean and variance, to vary too. Marginalising over these parameters permits a very wide range of bias scenarios to be considered, and greatly reduces the reliance on ad hoc assumptions about what the biases could look like. In this sense, our analysis is very conservative. In a later test (Section \ref{sec: multiple hypotheses}), however, we explicitly fix the bias mean and covariance to specific values in order to focus on the impact of different bias scenarios themselves. By giving the bias an uncertainty, we still allow for a range of bias values in this case however. 

\subsection{Early vs. Late Universe}
\label{sec: early vs late}

The most common approach to identifying the tension is by comparing early and late Universe measurements, which currently exhibit a 5$\sigma$ discrepancy \citep{Wang2023Hubble, Riess_2022Hubble, LIU2024138717}, making this the natural starting point for our analysis. In this work, we consider two definitions of the `early Universe': one where measurements with $z>2$ are classified as early, and a second, more inclusive case where all measurements with $z>0.2$ are taken as early. (This differs from some other commonly-used definitions, which might consider only $z \gtrsim 1000$ to be the early Universe.)

\begin{figure*}
    \centering
    \includegraphics[width=2\columnwidth]{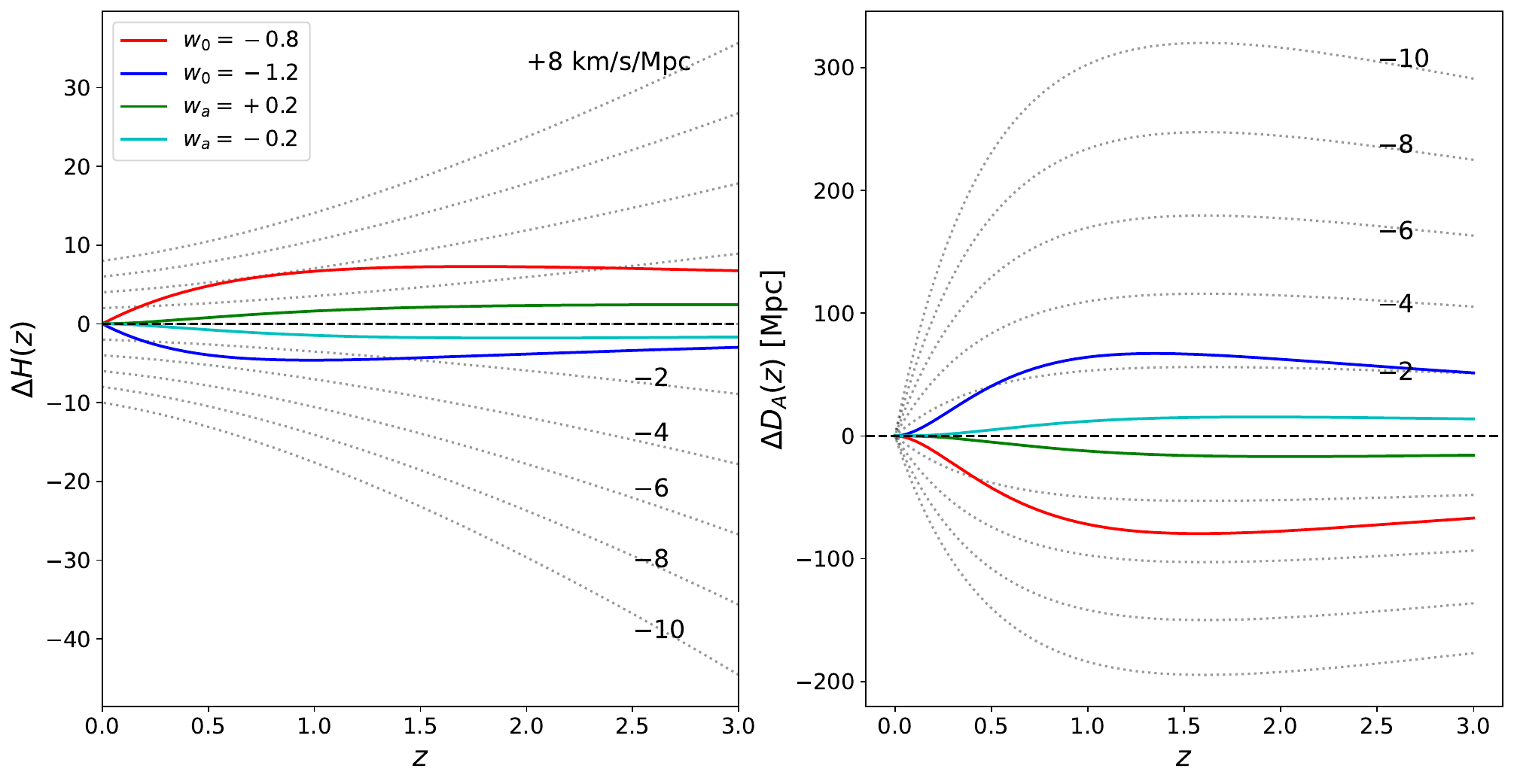}
    \caption{Differences in the cosmic expansion rate ($\Delta H(z)$) and angular diameter distance $D_A(z)$ relative to a reference $\Lambda$CDM cosmology (black dashed line), for dark energy models with the same value of $H(z=0)$. Coloured curves correspond to models with varying dark energy parameters ($w_0$, $w_a$), while keeping all other parameters fixed. Grey dotted lines show the effect of taking the $\Lambda$CDM model and shifting $H_0$ in increments of 2 \si{\km\per\second\per\mega\parsec}, providing a direct mapping between deviations in $H(z)$ at intermediate redshifts and their equivalent bias in the inferred $H_0$.}
    \label{fig:Delta H Cosmo}
\end{figure*}

In this comparison, we use subscripts on $\mathcal{M}$ to indicate which sets of measurements might be biased. Specifically, a model $\mathcal{M}_{X}$ represents a scenario in which the measurements labelled $X$ are biased, while all others remain unbiased. For example, $\mathcal{M}_{\text{E}}$ corresponds to a model in which only early Universe measurements are biased, $\mathcal{M}_{\text{L}}$ represents bias in late Universe measurements, and $\mathcal{M}_{\text{EL}}$ describes a case where both are biased. In general, $\mathcal{M}_{X_1 X_2...}$ indicates a bias affecting the measurements associated with $X_1, X_2, ...$, allowing us to flexibly represent different bias scenarios. Under each redshift cut, we test whether the tension arises from bias in either or both sets of measurements using four hypotheses: $\mathcal{M}_\text{E}$, $\mathcal{M}_\text{L}$, $\mathcal{M}_\text{EL}$, and $\mathcal{M}_\text{N}$. Here, $\mathcal{M}_\text{E}$ represents the hypothesis that only early Universe measurements are biased, $\mathcal{M}_\text{L}$ denotes that only late Universe measurements are biased, $\mathcal{M}_\text{EL}$ denotes that both sets are biased, and $\mathcal{M}_\text{N}$ denotes that none of the measurements are biased. The priors used in this test are
\begin{align}
    H_0 &\sim \mathcal{U}[60,80]~\si{\km\per\second\per\mega\parsec} \\
\label{bias mean}
    \mu_{\varepsilon} &\sim \mathcal{N}(0,10^2)~\si{\km\per\second\per\mega\parsec} \\
\label{bias variance}
    \sigma_{\varepsilon}^2 &\sim \mathcal{U}[0,10]~ \si{\km\squared\per\second\squared\per\mega\parsec\squared}.
\end{align}
Here, $\mathcal{U}(a, b)$ indicates a uniform probability distribution on the interval $(a,b)$, and $\mathcal{N}(\mu, \sigma^2)$ indicates a (univariate) Normal distribution with mean $\mu$ and variance $\sigma^2$. For this analysis, we adopted very broad, uninformative priors to ensure the results were driven primarily by the data rather than prior assumptions. The bias mean was assigned a wide normal prior centred at zero, reflecting no preferred direction for the bias but allowing for substantial deviations in either direction. The bias variance was given a uniform prior between 0 and 10, keeping it within reasonable bounds while still accommodating a wide range of possible systematic uncertainties. We assume that the bias covariance is diagonal, with constant values equal to this variance. The prior on $H_0$ was chosen to avoid unphysical extremes. Low values of $H_0$, such as $50$ \si{\km\per\second\per\mega\parsec}, would imply an unrealistically old Universe, while very high values, around $90$~\si{\km\per\second\per\mega\parsec}, would make it too young to fit well established astrophysical and cosmological constraints. We therefore restricted the prior to the range where most current measurements lie, excluding extreme values to improve computational efficiency and focus on the most relevant parameter space.

\subsection{Physical hypotheses}
\label{sec: multiple hypotheses}
To set up our physical hypotheses, we first need a consistent way of defining how different cosmological scenarios translate into biases on the inferred value of $H_0$. The central idea is that most high-redshift probes, such as the CMB or BAO, do not measure $H_0$ directly, but instead constrain the expansion rate $H(z)$ or distance measures at higher redshift. Inferring $H_0$ from these requires extrapolating within a cosmological model, so any deviation in $H(z)$ away from the reference $\Lambda$CDM expectation can propagate into an apparent bias in $H_0$. Ordinarily, this extrapolation would permit other cosmological parameters to shift too, but since our bias model only deals with $H_0$, we consider all other parameters to have been implicitly marginalised in a way that would have resulted in an acceptable $\Lambda$CDM model fit for each individual data point.

To illustrate this, consider a simple example at $z=2$ with a fixed reference cosmology. Suppose the observed value of $H(z=2)$ is larger than the $\Lambda$CDM prediction (based on the true value of $H_0$ at $z=0$) by 2 \si{\km\per\second\per\mega\parsec}, while all other cosmological parameters are fixed. We can use the Friedmann equation to write
\begin{equation}
H(z) = H_0 E(z); ~~~~~~~
E(z) = \sqrt{\Omega_m (1+z)^3 + \Omega_\Lambda}.
\end{equation}
The error on the inferred value of $H_0$ from that one data point is
\begin{equation}
\Delta H_0 = \frac{\Delta H(z)}{E(z)}.
\end{equation}
Using standard $\Lambda$CDM parameters $\Omega_m=0.3$ and $\Omega_\Lambda=0.7$, at $z=2$ we obtain
\begin{equation}
E(z=2) = \sqrt{0.3 \times 27 + 0.7} = \sqrt{8.8} \approx 2.97,
\end{equation}
which equates to an effective bias on $H_0$ of
\begin{equation}
\Delta H_0 = \frac{2}{2.97} \approx 0.67\, \si{\km\per\second\per\mega\parsec}.
\end{equation}
Thus, a $+2\, \si{\km\per\second\per\mega\parsec}$ shift in $H(z=2)$ corresponds to a positive offset of $\sim +0.7\, \si{\km\per\second\per\mega\parsec}$ in the extrapolated $H_0$ value, which will appear as a systematic bias when compared to unbiased direct measurements of $H_0$ in the current epoch.

Ideally, our chosen bias parameters for different physical scenarios would be calibrated by directly mapping deviations in the expansion history, $H(z)$, into effective biases in the inferred $H_0$. Figure \ref{fig:Delta H Cosmo} demonstrates this mapping, showing the differences in $H(z)$ across $z$ relative to a reference $\Lambda$CDM model. Coloured curves correspond to models with varying dark energy parameters ($w_0$, $w_a$), while dotted lines show the equivalent effect of shifting $H_0$ in increments of 2 \si{\km\per\second\per\mega\parsec}. This provides a direct quantitative link between physical changes to the expansion history and the bias terms introduced in our hypotheses. In practice it proved difficult to consistently match this mapping onto simple, interpretable scenarios however, and so for the main analysis we adopt a simplified set of bias specifications, based partly on these principles and partly on physical motivations such as LDL calibration errors or CMB modelling uncertainties. 

In contrast to the models in \S\ref{sec: early vs late}, the suite of physical hypotheses here does not easily allow for marginalisation over the bias mean and variance. Marginalising over bias parameters in these cases would blur the distinctions between physically motivated models, making their interpretations ambiguous and preventing meaningful comparison between them. Instead, we treat each physical hypothesis as a complete model with a fixed structure, and compute its posterior probability using Bayes' theorem (Equation \ref{eq:bayes_theorem}). This approach enables consistent model comparison within a framework where each hypothesis corresponds to a unique, physically motivated bias scenario.

\begin{figure}
    \centering
    \includegraphics[width=\columnwidth]{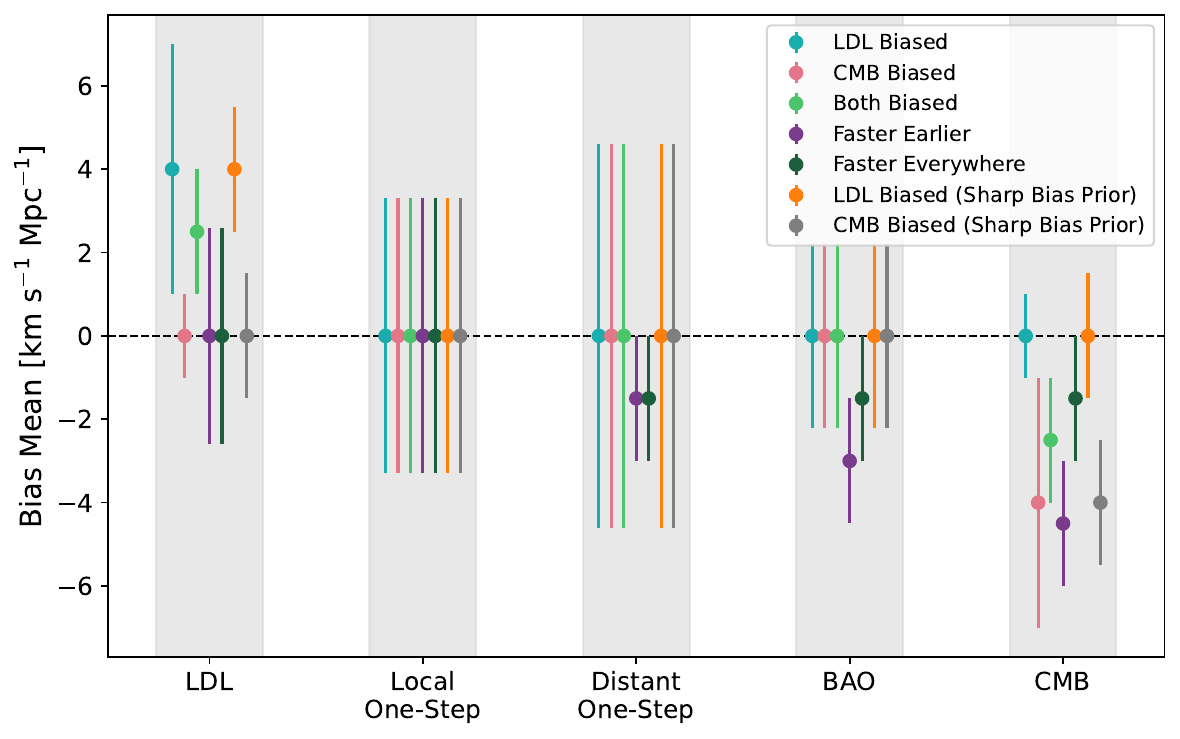}
    \caption{Bias specifications used in the analysis (as shown in Table \ref{tab: bias specs}, grouped by measurement class: LDL, Local One-step, Distant One-step, BAO, and CMB. Each point represents the bias mean $\mu_\varepsilon$, with error bars corresponding to the bias uncertainty. For parameters expressed in terms of the reported measurement uncertainties ($\sigma$) in table, the plotted value corresponds to the median of the reported uncertainty range. While the plot shows only these representative medians for clarity, the full analysis used the complete set of reported measurements and their uncertainties.}
    \label{fig:bias specs}
\end{figure}

There are eight total hypotheses we compare in this case: `LDL scenario', `CMB scenario', `Both Biased', `Faster Earlier', `Faster Everywhere scenario', `LDL Biased (Sharp Bias Prior)', and `CMB Biased (Sharp Bias Prior)'. Each model $\mathcal{M}_j$ in this set corresponds to a fully specified structure with fixed values for the bias mean and bias covariance, reflecting a physical interpretation (e.g., specific systematic effects). Bias specifications for each scenario are summarised in Table \ref{tab: bias specs} and shown in Figure \ref{fig:bias specs}. 

In the LDL scenario, LDL measurements are assumed to be biased high due to calibration systematics \citep{Scolnic2024, LIU2024138717}.
The CMB scenario is motivated by the recognition that systematic effects or modelling assumptions in the analysis of CMB data could lead to a downward bias in the inferred value of $H_0$ \citep{Weikang2017Discordance, Poulin_2019, DiValentino2021, yeung2024resolvingh0s8tensions, Aboubrahim_2024, Kou2024, Lynch2024, Clifton_2024}. The `both biased' scenario then assumes that both LDL and CMB measurements are biased by a similar amount but in opposite directions, which would mean the true value of $H_0$ lies somewhere in-between both estimates. 

The bias covariance structure in these cases is taken to be diagonal, i.e. we treat the bias terms for each measurement as independent. This choice reflects the fact that the LDL, CMB, and other probes are derived from very different physical observables, operating at widely separated redshifts and relying on distinct analysis pipelines. Due to this separation, it is unlikely that the same systematic error would coherently affect multiple categories in a correlated way. While correlations within a class are possible, a hypothesis that does not have correlations helps us probe situations that cause excess scatter in a class beyond the quoted statistical uncertainties, which may be useful for determining the root cause of the bias. In particular, setting intra-class correlations to zero describes an additional experimental error that varies randomly from experiment-to-experiment (even within a given class of experiments), rather than a bias whose value is common to all experiments of that class. 

\begin{table}
    \centering
    \setlength{\tabcolsep}{2pt}
    \begin{tabular}{llcc}
        \toprule
        \textbf{Scenario} & \textbf{Measurement} & \textbf{Bias Mean} & \textbf{Bias Std. Dev} \\
        \textbf{} & \textbf{} & \textbf{(\si{\km\per\second\per\mega\parsec})} & \textbf{$\pm$ (\si{\km\per\second\per\mega\parsec})} \\
        \midrule \midrule
        LDL Biased & LDL & $+4$ & $3$ \\
        & Local One-Step & $0$ & $\sigma$ \\
        & Distant One-Step & $0$ & $\sigma$ \\
        & BAO & $0$ & $\sigma$ \\
        & CMB & $0$ & $1$ \\        
        \hline
        CMB Biased & LDL & $0$ & $1$ \\
        & Local One-Step & $0$ & $\sigma$ \\
        & Distant One-Step & $0$ & $\sigma$ \\
        & BAO & $0$ & $\sigma$ \\
        & CMB & $-4$ & $3$ \\        
        \hline
        Both Biased & LDL & $+2.5$ & 1.5 \\
        & Local One-Step & $0$ & $\sigma$ \\
        & Distant One-Step & $0$ & $\sigma$ \\
        & BAO & $0$ & $\sigma$ \\
        & CMB & $-2.5$ & 1.5 \\ 
        \hline
        Faster Earlier & LDL & $0$ & $\sigma$ \\
        & Local One-Step & $0$ & $\sigma$ \\
        & Distant One-Step & $-1.5$ & 1.5 \\
        & BAO & $-3$ & $1.5$ \\
        & CMB & $-4.5$ & $1.5$ \\
        \hline
        Faster Everywhere & LDL & $0$ & $\sigma$ \\
        & Local One-Step & $0$ & $\sigma$ \\
        & Distant One-Step & $-1.5$ & 1.5 \\
        & BAO & $-1.5$ & $1.5$ \\
        & CMB & $-1.5$ & $1.5$ \\
        \hline
        LDL Biased & LDL & $+4$ & $1.5$ \\
        (Sharp Bias Prior) & Local One-Step & $0$ & $\sigma$ \\
        & Distant One-Step & $0$ & $\sigma$ \\
        & BAO & $0$ & $\sigma$ \\
        & CMB & $0$ & $1$ \\        
        \hline
        CMB Biased & LDL & $0$ & $1$ \\
        (Sharp Bias Prior) & Local One-Step & $0$ & $\sigma$ \\
        & Distant One-Step & $0$ & $\sigma$ \\
        & BAO & $0$ & $\sigma$ \\
        & CMB & $-4$ & $1.5$ \\        
        \bottomrule
    \end{tabular}
    \caption{Bias specifications and standard deviations of measurements for the quantitative scenarios. Here, $\sigma$ is the reported error bar of the associated measurement.}
    \label{tab: bias specs}
\end{table}

We also define `Sharp Bias Prior' variants of the LDL and CMB biased scenarios. These are identical to their original counterparts except that the standard deviation on the relevant subset is reduced from 3 to 1.5. This reflects greater confidence in the magnitude of the bias within that data category. The reduced bias variance implies a narrower range of plausible bias values, and therefore smaller amounts of excess scatter in the observed data. 

Finally, we construct two physically motivated scenarios reflecting modified expansion histories. These scenarios are constructed to produce a bias of similar magnitude to the CMB and LDL biased scenarios, since they are designed to serve as potential solutions to the Hubble tension. Any viable resolution should lead to a similar overall effect, just through different observables or redshift combinations. In the `Faster Earlier' case, the expansion rate is systematically higher at earlier times, corresponding to bias means of $(-4.5, -3, -1.5)$ \si{\km\per\second\per\mega\parsec} for CMB, BAO, and distant one-step respectively. This gives a trend of decreasing bias from high to intermediate redshift.

In the `Faster Everywhere' case, the expansion is uniformly higher at $z>0$, with a common bias of $-1.5$ across distant probes, while LDL and local one-step remain unbiased. Both scenarios assume a standard deviation of 1.5. As these represent coherent shifts at similar redshifts, they are implemented with a block-diagonal covariance structure, assuming perfect correlations within each measurement group. For example, all LDL measurements are treated as fully correlated with each other, as are all CMB measurements etc. Although some degree of cross-redshift correlation would be expected in a fully specified model, assigning these would require committing to a particular physical mechanism. To avoid making such assumptions, we conservatively restrict correlations to within each redshift block.

Clearly, many other possible bias model specifications are available, and important aspects such as consistency with multi-probe measurements of other cosmological parameters have been omitted. Our goal here is only to span a broad enough range of interesting scenarios, without unduly favouring particular types of hypothesis or regions of the available model space. Nevertheless, our code is publicly-available\footnote{\href{https://github.com/QuanTom97/Assessment-of-the-Hubble-tension-via-Bayesian-jackknife-testing}{https://github.com/QuanTom97/Assessment-of-the-Hubble-tension-via-Bayesian-jackknife-testing}} so that interested readers can try their own bias hypothesis specifications.

\section{Results}
\label{sec: results}

In this section, we present results for the early and late Universe comparison, for each definition of `early Universe'. We then show results for the broader set of phenomenological hypotheses and then combine the results across these models to obtain a marginalised posterior for $H_0$. 

\subsection{Early vs. Late Universe}
\label{sec:early_vs_late_results}

\begin{figure}
    \centering
    \includegraphics[width=\columnwidth]{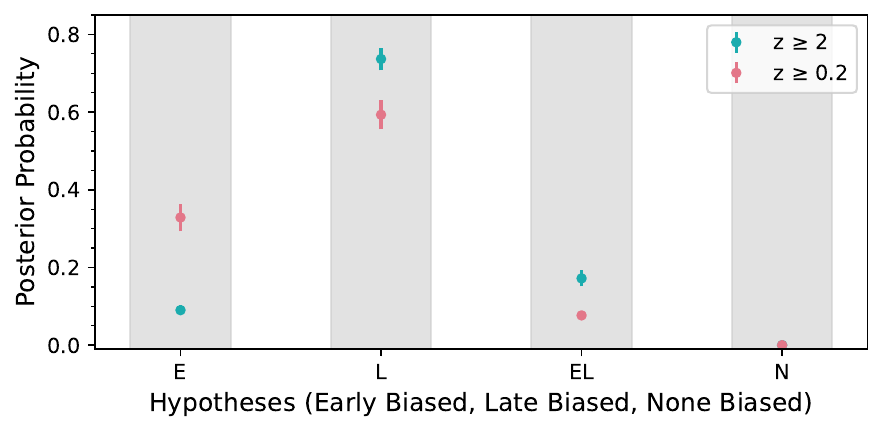}
    \caption{Posterior support for early versus late bias scenarios under two definitions of the early/late division. The $z \geq 2$ split (shown in blue) isolates the CMB as early, with all other probes treated as late, while the $z \geq 0.2$ split (shown in red) assigns local probes (LDL and low-$z$ one-step) to the early category and higher-redshift probes to the late category. In both cases, late bias is favoured, though the strength of this preference depends on the placement of the cut-off.}
    \label{fig:early_vs_late_16_dp_sum}
\end{figure}

Figure \ref{fig:early_vs_late_16_dp_sum} shows the posterior support for both versions of the early versus late analysis, computed using Bayes' theorem (Equation \ref{eq:bayes_theorem}), assuming a flat prior over models. In the $z \geq 2$ split, the results show a strong preference for the late biased hypothesis ($\mathcal{M}_\text{L}$), supported at close to the $\sim 0.8$ relative probability level. The next most supported hypotheses are the combined early and late case ($\mathcal{M}_\text{EL}$) and the purely early biased case ($\mathcal{M}_\text{E}$), both at the $\sim 0.15 - 0.2$ level. This corresponds to odds of roughly 4:1 in favour of $\mathcal{M}_\text{L}$ over the next most likely hypothesis. The unbiased case ($\mathcal{M}_\text{N}$) again has negligible support -- there is definitely a tension between the early and late measurements when they are grouped in this way.

This preference could be partly explained by the relative composition of the two categories. The early Universe group consists of only two measurements (CMB), which are internally consistent and tightly constrained, leaving little room for them to drive strong tensions. By contrast, the late Universe category contains 14 measurements across LDL and one-step methods, which show a wider spread in central values and uncertainties. In other words, the results of the test demonstrate that the early Universe measurements are internally consistent, making them less likely to be identified as biased. The posterior support for the combined bias case ($\mathcal{M}_\text{EL}$) also reflects this, as the spread among late-time measurements makes it plausible that both early and late measurements could contribute to the tension, though not as strongly as the late-only scenario.

Ultimately, the larger number and greater variability of late Universe measurements help explain why $\mathcal{M}_\text{L}$ emerges as the preferred hypothesis. Crucially, the early Universe measurements have smaller error bars, meaning that any excess variation would be immediately apparent, making them especially vulnerable to being flagged as biased. The fact that $\mathcal{M}_\text{L}$ is still favoured, despite this, provides stronger evidence against the late measurements. Rather than ruling out early Universe bias, the result highlights the relative internal consistency of the early data compared to the broader and more scattered late-time observations.

In the $z \geq 0.2$ split, the most favoured hypothesis is again the late biased case ($\mathcal{M}_\text{L}$), with posterior probability of around $0.6$. However, the early biased hypothesis ($\mathcal{M}_\text{E}$) also retains substantial support at $\sim 0.35 - 0.4$, giving odds of only $\sim$ 1.5:1 in favour of $\mathcal{M}_\text{L}$ over $\mathcal{M}_\text{E}$. The joint bias scenario ($\mathcal{M}_\text{EL}$) receives little support, while the unbiased case ($\mathcal{M}_\text{N}$) is again negligible. This reduced contrast compared to the $z \geq 2$ split arises because the broader cut includes several one-step measurements in the early category. Many of these have comparatively large uncertainties, such as data point 11 (gravitational waves, $H_0 = 64.67^{+6.9}_{-10.7}$ \si{\km\per\second\per\mega\parsec}) and central values that lie closer to the CMB estimate ($H_0 \approx 68$ \si{\km\per\second\per\mega\parsec}). When combined with the precise CMB measurements, this makes it appear less likely that the early Universe measurements are driving the tension and shifts part of the posterior weight toward $\mathcal{M}_\text{L}$.

Within the one-step set, three particularly low values, data points 12 (FRBs, $H_0 = 64.7 ^{+5.6}_{-4.7}$ \si{\km\per\second\per\mega\parsec}), 14 (gamma-ray attenuation, $H_0 = 62.4 ^{+4.1}_{-3.9}$ \si{\km\per\second\per\mega\parsec}), and 15 (time-delay cosmography, $H_0 = 66 \pm 4$ \si{\km\per\second\per\mega\parsec}), further pull the inference toward the CMB values. Meanwhile, outliers in the LDL measurements also contribute: for example, data point 6 (JAGB-SNIa, $H_0 = 68 \pm 2.7$ \si{\km\per\second\per\mega\parsec}) sits close to the CMB, while data point 7 (Tully–Fisher, $H_0 = 76 \pm 2.6$ \si{\km\per\second\per\mega\parsec}) is notably high compared to the rest of the LDL data, amplifying internal inconsistency within the LDL set. Taken together, these features help explain why the analysis still favours late Universe bias overall, but with a less decisive preference than in the $z \geq 2$ case, leaving open the possibility that systematic effects in low-to-intermediate redshift probes could also play a role.

\subsection{Physical Hypotheses}
\label{sec: multiple hyp results}

\begin{figure*}
    \centering
    \includegraphics[width=\textwidth]{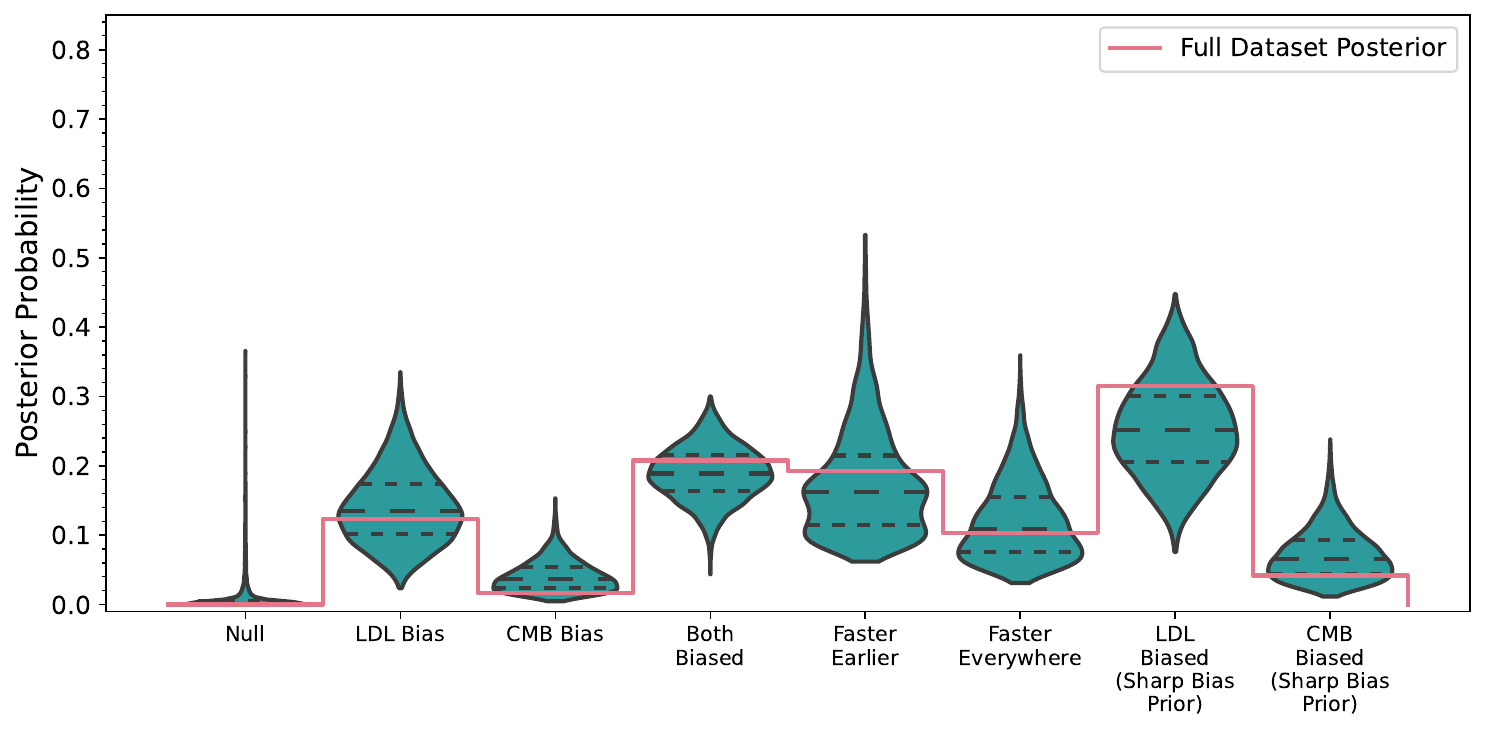}
    \caption{Violin plot illustrating the posterior distributions for each hypothesis in the multiple hypothesis case over the ensemble of N=11 subsets of the N=16 original data points. The width of each violin represents the density of the posterior probabilities. Dashed lines within each violin indicate the median and interquartile (25$\%$ and 75$\%$) ranges. Models such as Faster Earlier, LDL Biased, and LDL Biased (Sharp Bias Prior) receive higher support compared to alternatives, while the null and CMB biased scenarios are consistently disfavoured. The inference overall is uncertain on the true source of the tension, with no singular model emerging as dominant compared to the others. The red step lines show posterior weights computed from the full 16-point dataset. Models such as Both Biased, LDL Biased (Sharp Bias Prior), and Faster Earlier receive the most consistent support across subsets, whereas null and CMB bias hypotheses are disfavoured. However, no single scenario dominates decisively, highlighting continued uncertainty about the true origin of the Hubble tension.} 
    \label{fig:violin plot}
\end{figure*}

In this section, we compare the posterior probabilities for the hypotheses outlined in Section \ref{sec: multiple hypotheses}. While the goal is to evaluate the relative support for each scenario, we also recognise that the specific selection of data points can influence the outcome. To assess this sensitivity, we performed a resampling analysis, a method commonly used to test the robustness of statistical inferences to the inclusion or exclusion of individual measurements. Specifically, we generated all possible subsets of 11 measurements from the complete set of 16 $H_0$ values, resulting in 4,368 unique combinations. This allowed us to systematically examine how support for different bias hypotheses changes with different data configurations. The analysis was carried out using \texttt{CHIBORG}. 

Figure \ref{fig:violin plot} presents the posterior probabilities for each bias scenario, computed over all $N=11$ jackknife subsets of the original 16-point dataset. Each `violin' represents the distribution of posterior weights across these subsets, with wider regions corresponding to higher density. The red step line overlays the posterior computed from the full dataset. This shows that the LDL Biased (Sharp Bias Prior), Both Biased, and Faster Earlier scenarios receive the greatest support when all data are included. In contrast, the CMB Biased and Null scenarios are consistently disfavoured, both in the full-data posterior and across nearly all jackknife subsets. The violin distributions illustrate that while individual subset posteriors can fluctuate, the main inference is robust: support remains concentrated among the same small group of models even when measurements are removed.

\begin{figure*}
    \centering
    \includegraphics[width=1.9\columnwidth]{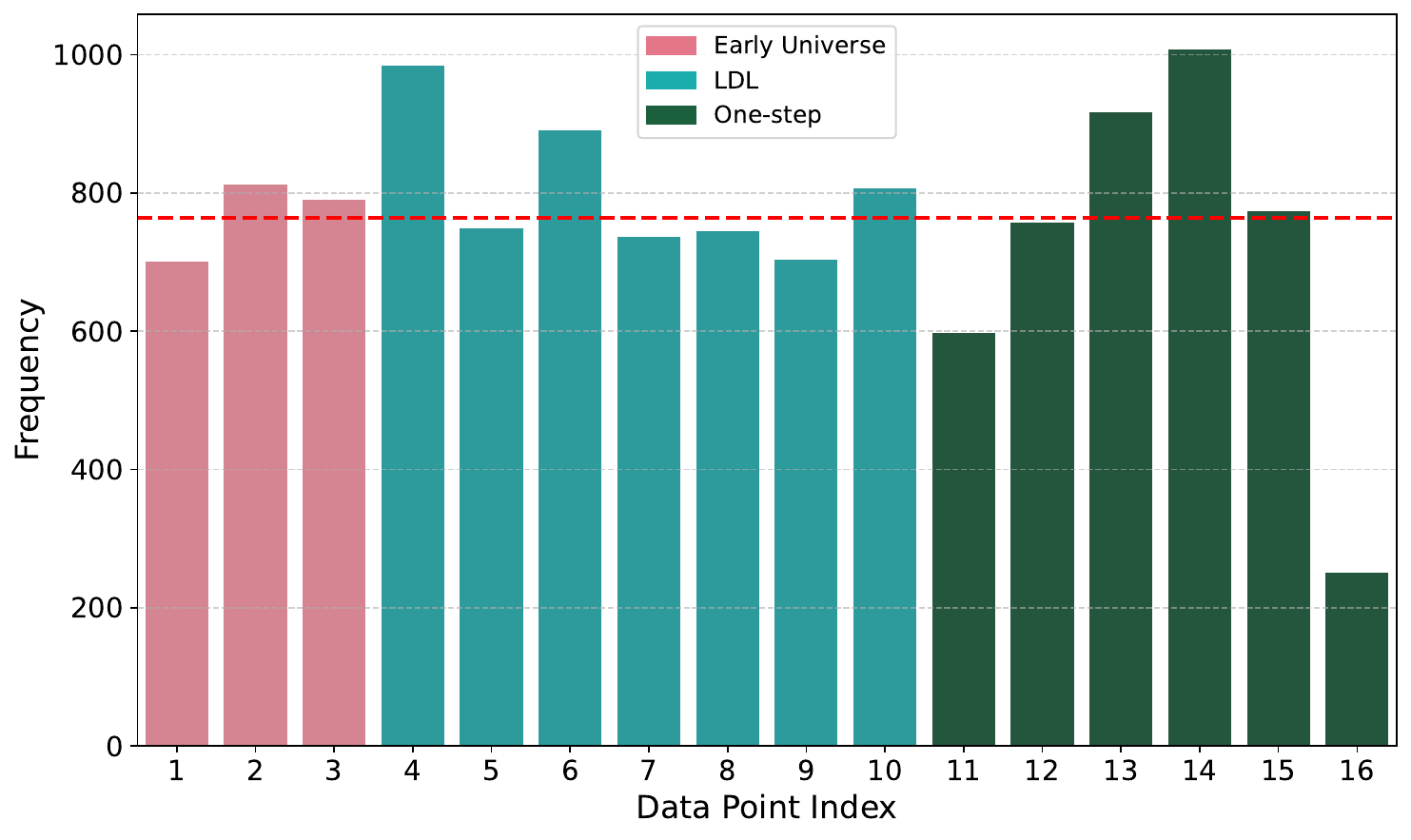}
    \caption{Frequency of data points appearing in subsets where the posterior probability for the LDL Biased (Sharp Bias Prior) hypothesis exceeds 0.3. Bars are coloured by data category. A total of 1,111 such cases are included. The red-dashed line marks the null frequency of 764, which corresponds to the expected count if these 1,111 cases were randomly chosen from the original 4,368 subsets with uniform probability. Data point 16 ($H_0 = 73.9 \pm 3$ \si{\km\per\second\per\mega\parsec}; masers) appears notably less often than other measurements, indicating that its exclusion is frequently associated with stronger support for the LDL Biased (Sharp Bias Prior) scenario. Similarly, data point 11 ($H_0 = 64.67^{+6.9}_{-10.7}$ \si{\km\per\second\per\mega\parsec}; gravitational waves) appears with lower frequency, while points ($H_0 = 73.2 \pm 1.3$ \si{\km\per\second\per\mega\parsec}; Cepheids-SNIa), 6 ($H_0 = 68 \pm 2.7$ \si{\km\per\second\per\mega\parsec}; JAGB-SNIa), 13 ($H_0 = 68 \pm 3.6$ \si{\km\per\second\per\mega\parsec}; EPM + Kilonovae), and 14 ($H_0 = 62.4 ^{+4.1}_{-3.9}$ \si{\km\per\second\per\mega\parsec}; gamma-ray attenuation) appear more often, suggesting their inclusion is associated with higher posterior support for LDL bias.} 
    \label{fig:new data hist}
\end{figure*}

The medians of the distributions are often roughly consistent with the full-data posterior, though there is one notable shift, in the LDL Biased (Sharp Bias Prior) scenario, where the median sits noticeably below the red line.  Further inspection reveals that including data points 4 ($H_0 = 73.2 \pm 1.3$ \si{\km\per\second\per\mega\parsec}; Cepheids-SNIa), 6 ($H_0 = 68 \pm 2.7$ \si{\km\per\second\per\mega\parsec}; JAGB-SNIa), 13 ($H_0 = 68 \pm 3.6$ \si{\km\per\second\per\mega\parsec}; EPM + Kilonovae), and 14 ($H_0 = 62.4 ^{+4.1}_{-3.9}$ \si{\km\per\second\per\mega\parsec}; gamma-ray attenuation), and excluding 11 ($H_0 = 64.67^{+6.9}_{-10.7}$ \si{\km\per\second\per\mega\parsec}; gravitational waves) and 16 ($H_0 = 73.9 \pm 3$ \si{\km\per\second\per\mega\parsec}; masers), tends to produce posterior values for LDL Biased (Sharp Bias Prior) above 0.3 (see Figure \ref{fig:new data hist}), which is roughly the median of its distribution.

This pattern is most likely driven by the position of these measurements relative to the LDL estimates. For instance, data point 14 ($H_0 = 62.4 ^{+4.1}_{-3.9}$ \si{\km\per\second\per\mega\parsec}; gamma-ray attenuation) is one of the lowest $H_0$ measurements in the dataset, making it strongly supportive of a model where the LDL measurements are biased high. Similarly, data point 16 ($H_0 = 73.9 \pm 3$ \si{\km\per\second\per\mega\parsec}; masers) is a one-step measurement that sits near the centre of the LDL distribution, so this would naturally support LDL measurements being biased when excluded. This particular measurement provides strong support for the LDL measurements being unbiased. Its inclusion consistently increases posterior support for models in which LDL is treated as unbiased, so removing it effectively eliminates one of the key contributors supporting an unbiased interpretation of the LDL data.

Overall, multiple hypotheses receive comparable posterior weight, with no single explanation emerging as decisively favoured. The jackknife subset selection shows that this conclusion is quite robust to the choice of data, with only the Faster Earlier hypothesis exhibiting a long tail that suggests some sensitivity to this choice. The LDL Biased (Sharp Bias Prior) scenario most often has the highest probability, but does not reach even a 2:1 odds ratio with the Both Biased and Faster Earlier scenarios on average, making it only weakly favoured. In addition to the null hypothesis, a couple of scenarios are generically disfavoured however -- these are the two CMB Bias scenarios, although even then there are a few data selections that would yield posterior probabilities of $0.15 - 0.2$ for these hypotheses. 

\subsection{Estimating the Hubble parameter posterior distribution}

We next estimate the posterior distribution of $H_0$ by marginalising over the available models. To explore this, we set up a one-parameter model in \texttt{PolyChord} for each hypothesis, where the only free parameter is $H_0$. The bias term $\varepsilon$ is analytically marginalised over by fixing the bias parameters (bias mean and covariance) in advance (Equation \ref{eq:fix_like}). As a result, the posterior under each hypothesis reduces to a function of $H_0$ alone,
\begin{equation}
    P(H_0|\textbf{d}, \mathcal{M}_j) \propto P(\textbf{d}|H_0, \mathcal{M}_j)P(H_0).
\end{equation}
We can then combine these posteriors into a single marginalised distribution by weighting each one according to the evidence for its respective model (again, assuming a flat prior over models). From Bayes' theorem, 
\begin{equation}
    P(H_0|\textbf{d})=\sum_j P(H_0,\mathcal{M}_j|\textbf{d})
\end{equation}
Then applying the product rule, this becomes:
\begin{equation}
    P(H_0|\textbf{d})=\sum_j P(H_0|\mathcal{M}_j,\textbf{d})P(\mathcal{M}_j|\textbf{d})
\end{equation}
Here, $P(H_0|\mathcal{M}_j,\textbf{d})$ is the posterior for the parameters under model $\mathcal{M}_j$, which \texttt{PolyChord} samples. The term $P(\mathcal{M}_j|\textbf{d})$ represents the posterior probability of each model, which is proportional to the Bayesian evidence also calculated by \texttt{PolyChord}. The result is a single, model-weighted posterior distribution for $H_0$.

\begin{figure*}
    \centering
    \includegraphics[width=\columnwidth]{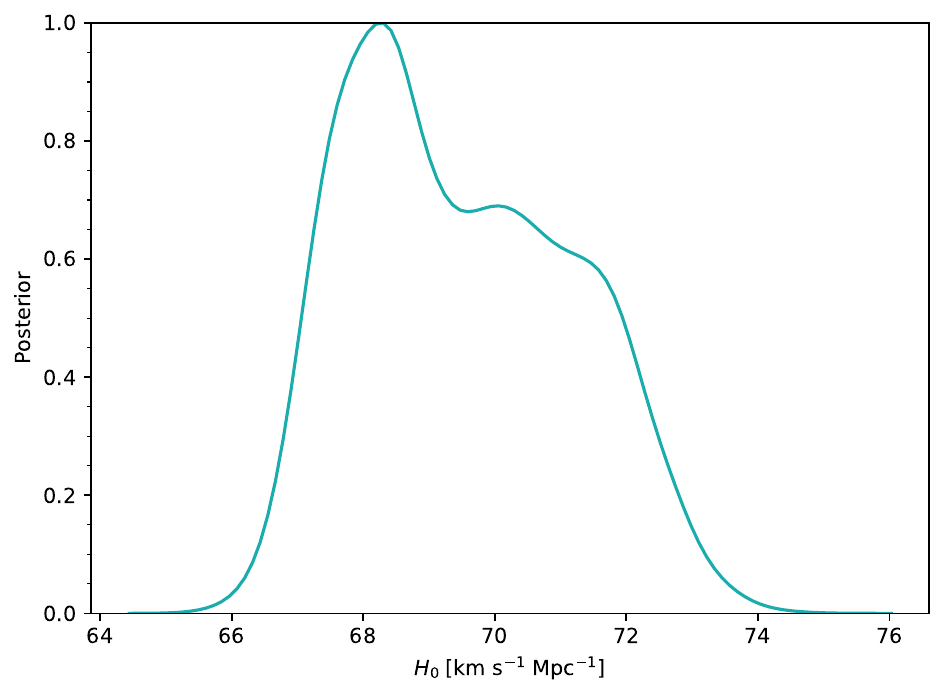}
    \includegraphics[width=\columnwidth]{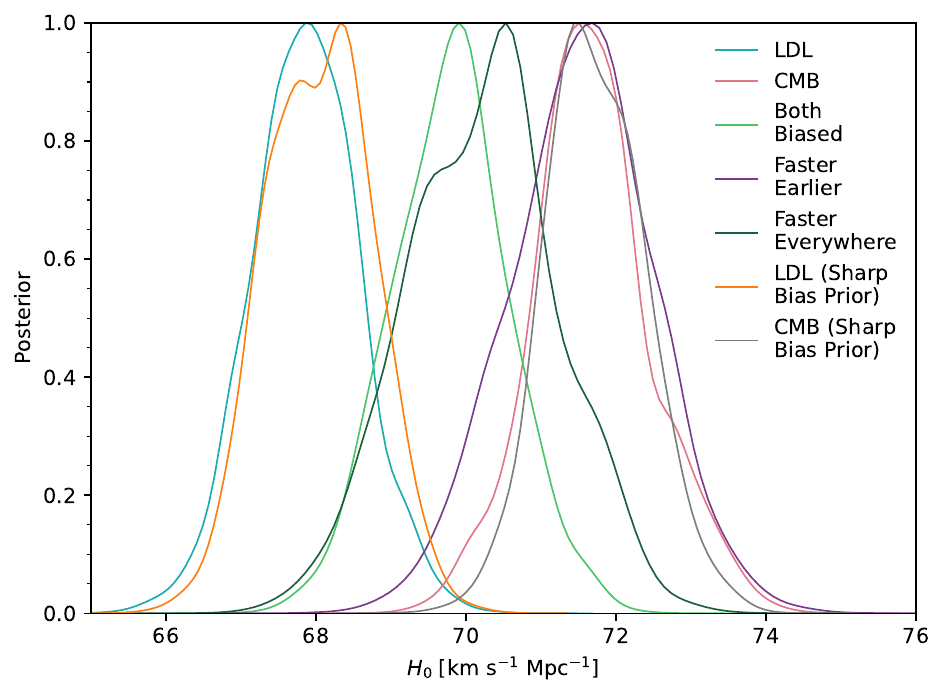}
    \caption{(Left): Posterior distribution of $H_0$ after marginalising over all bias models, with each weighted by its Bayesian evidence. The resulting distribution is notably broad and exhibits three distinct peaks, reflecting the ambiguity in the underlying source of the tension. The main peak near $H_0 \sim 68$ \si{\km\per\second\per\mega\parsec} likely reflects the influence of precise CMB measurements, while the secondary peaks correspond to broader support from other bias scenarios and measurement subsets. This highlights the uncertainty and model-dependence in resolving the $H_0$ tension. (Right): The posteriors for the individual models.}
    \label{fig:marginalised H0}
\end{figure*}

Figure \ref{fig:marginalised H0} presents the marginal posterior on $H_0$ after marginalising over the bias models. The posterior here is notably broad and ambiguous, with an almost trimodal structure. This inference finds the value of $H_0$ -- assuming that there is a phenomenological explanation for the Hubble tension -- to be in the range $66.7 < H_0 < 72.7$ \si{\km\per\second\per\mega\parsec} (95$\%$ credible region). The maximum a posteriori value is around 68~\si{\km\per\second\per\mega\parsec}, likely reflecting the small measurement uncertainties on the CMB measurements. This peak is far from dominant however, with substantial weight also appearing around $H_0 \approx 70$ and $H_0 \approx 71.5$ \si{\km\per\second\per\mega\parsec}. These additional peaks are consistent with support for non-trivial hypotheses in the earlier analysis, particularly the `Both Biased' and `Faster Earlier' scenarios, as well as possible contributions from wider error bars in LDL or one-step measurements.

The individual marginal posterior distributions for $H_0$ are shown in the right panel of Fig.~\ref{fig:marginalised H0}. None of the hypotheses have disjoint support; they all overlap to some extent. For the LDL and CMB hypotheses, the shape of the posterior remains relatively stable when stronger priors are used (the `Sharp Bias Prior' versions of these hypotheses). The `Both Biased' hypothesis does indeed produce a marginal value of $H_0$ that sits between the LDL and CMB hypothesis values. Comparing with the fully marginalised distribution in the left panel, however, it can be seen that the LDL scenario is the most influential; just not sufficiently to seriously rule-out models in which a higher `true' value of $H_0 \gtrsim 70$~\si{\km\per\second\per\mega\parsec} is preferred.

The marginal posterior shown in Fig.~\ref{fig:marginalised H0} (left panel) could potentially be used as a `model agnostic' best-estimate of $H_0$. This would be useful for analyses that do not seek to explain the Hubble tension, and only need a good estimate of the value of $H_0$ and its corresponding uncertainty in light of the fact that the tension does exist. This posterior distribution was calculated by including a wide range of the best available measurements, and allowed for a variety of explanations for the observed tension, without committing to any given one. The Bayesian evidence for each scenario was used to weight its contribution to the full marginalised posterior, rather than a manually-chosen ad hoc prior. We also verified that there is a reasonable level of insensitivity to the choice of priors (e.g. in comparing with the `Sharp Bias Prior' versions of the LDL and CMB hypotheses) and the choice of data points. While additional hypotheses with different bias specifications could be added, or more sophisticated model and nuisance parameter treatments could be used, we would not expect these results to change radically unless a highly specific and explanatory hypothesis for the cause of the tension is found.


\section{Summary and Conclusions}
\label{sec: conclusion}

In this paper, we systematically examined possible explanations of the Hubble tension by comparing early and late Universe measurements, with a couple of different ways to define the early Universe measurements. We then evaluated a broad set of alternative phenomenological scenarios. Using our extended \texttt{CHIBORG} code and \texttt{PolyChord}, we conducted a robust Bayesian analysis and obtained marginalised estimates of $H_0$.

In our early-late Universe jackknife analysis, our results show that the inference depends on how the early-late division is defined, but in both cases the data favour the late Universe measurements as the more likely source of bias. With a strict cut at $z \geq 2$, the late Universe biased hypothesis favours the next best alternative by 4:1, while with the broader cut at $z \geq 0.2$ the preference is weaker (1.5:1), with substantial support also remaining for the early Universe biased hypothesis. This reduced contrast may arise because the $z \geq 0.2$ cut includes several one-step measurements in the early category, many of which have larger error bars and central values closer to the CMB, diluting the observed distinction between the early and late subsets. Additionally, influential LDL outliers amplify internal inconsistencies within the late subset. These results indicate that while generic late Universe systematics remain a plausible explanation of the tension, the strength of this conclusion is sensitive to the chosen redshift split.

In addition to the early-late comparison, we tested a broader suite of phenomenological bias scenarios, incorporating both physically motivated and more agnostic constructions. To assess robustness, we carried out a jackknife resampling over all $\genfrac(){0pt}{2}{16}{11}$ subsets of the data, generating posterior distributions for each hypothesis. This analysis revealed that, while no single scenario dominates across all subsets, the full 16-point dataset consistently favours a small group of models, most prominently the LDL Biased (Sharp Bias Prior), Faster Earlier, and Both Biased hypotheses. The Null and CMB Biased scenarios remain strongly disfavoured in both the full-data posterior and nearly all resamples. The jackknife distributions demonstrate that the inference remains broadly consistent across different data subsets, with median posterior weights typically aligning closely with the full-data result. However, some shifts do occur when specific influential measurements are removed, highlighting limited but informative sensitivity to dataset composition. Nevertheless, the full-data posterior emerges as the clearest indicator of model preference, reinforcing the interpretation that late-time measurements, particularly within the LDL category, are the more likely drivers of the observed tension, although this preference is only mild.

We estimated $H_0$ by marginalising over the alternative phenomenological scenarios and combined them, applying a weighting based on the Bayesian evidence-based probability of each scenario. The distribution took on a multi-modal shape, with a maximum a posterior value of roughly $68$~\si{\km\per\second\per\mega\parsec}, within a range of $66.7 < H_0 < 72.7$ \si{\km\per\second\per\mega\parsec} (95$\%$ credible region). This value is likely driven by the tighter CMB error bars. The broader structure of the posterior distribution reflects a greater ambiguity in the true value of $H_0$, consistent with the distributions of the violin plots, which show that no single hypothesis receives dominant support. The small middle peak appears to be primarily driven by the `Faster Everywhere' scenario, which consistently favours a value of $H_0 \approx 70$ \si{\km\per\second\per\mega\parsec}. This is similar to the result found in the `Both Biased' case, where the true value lies between the early and late Universe estimates. Ultimately, this shift in the posterior highlights the extent to which our conclusions depend on how the models are defined.

The robustness of these results is supported by several aspects of our approach. First, the bias mean and bias variance were treated as free parameters with flat priors in the early versus late Universe tests, allowing for a full exploration of plausible bias configurations. Second, in the context of the phenomenological scenario analysis, the use of resampling over all 11 point subsets of the 16 data points allowed us to map the sensitivity of the conclusions to the chosen dataset. By comparing these subset-based posteriors to the results obtained using the full dataset, we confirmed that the main conclusions remain broadly stable. While individual subset results varied, support consistently clustered around the same small group of models, with the full-data posterior typically falling near the median of the subset distributions. The only notable deviation was seen in the LDL Biased (Sharp Bias Prior) scenario, where the full data posterior sat noticeably above the median, suggesting that a small number of data points may disproportionately influence its support.

Looking ahead, resolving the Hubble tension will require either new, independent measurements, or a deeper understanding of systematic effects in existing data, so that they can be modelled more explicitly. Given the challenge of acquiring more truly independent $H_0$ measurements, a promising path forward is to more accurately model correlations between non-independent measurements. Once correlations are well characterised, the same hypothesis, marginalisation and sensitivity techniques used here can be applied to a more refined, larger dataset. Finally, while our bias specifications were physically motivated, an improved approach would be to calibrate them by mapping deviations in $H(z)$ and related distance measures directly into effective $H_0$ biases, or by carrying out detailed self-consistent analyses within each scenario. We outlined how this can be done in principle, but a full implementation would require detailed modelling of the expansion history across redshift, involving multiple cosmological parameters and assumptions, potentially for very many plausible models. As a starting point, however, incorporating such a model in future work would provide a more physically motivated connection between theory and the apparent biases that may be shifting $H_0$ measurements away from the true value.

\section*{Acknowledgements}

We are grateful to R.~Battye, J.~Chluba, S.~Dhawan, and F.~Kennedy for useful discussions. We acknowledge the use of \texttt{PolyChord} \citep{Handley2015a, Handley2015b}, \texttt{anesthetic} \citep{Handley2019}, and \texttt{CHIBORG} \citep{Wilensky2022Bayesian}, in carrying out this analysis. This result is part of a project that has received funding from the European Research Council (ERC) under the European Union’s Horizon 2020 research and innovation programme (Grant agreement No. 948764).

\section*{Data Availability}

The code and data used to perform the Bayesian jackknife tests, including those used to generate key plots and results in this work, are available from GitHub.\footnote{\href{https://github.com/QuanTom97/Assessment-of-the-Hubble-tension-via-Bayesian-jackknife-testing}{https://github.com/QuanTom97/Assessment-of-the-Hubble-tension-via-Bayesian-jackknife-testing}} The data used in this analysis are already publicly available; we took the first 11 measurements from Table \ref{tab:16 Hubble measurements} from an online repository\footnote{\url{https://github.com/lucavisinelli/H0TensionRealm/blob/main/data/dataset.csv}}, while the next five measurements were taken from published articles, as referenced in Table \ref{tab:16 Hubble measurements}.



\bibliographystyle{mnras}
\bibliography{hubble_tension_bayesian} 




\appendix

\section{Analytic Solution with Gaussian Priors}
\label{Appendix A}

In this appendix, we outline the mathematical formalism used to incorporate Gaussian priors into this analysis. This process follows the same approach outlined in \citep{Wilensky2022Bayesian}. We define the data vector $\mathbf{d}$ as

\begin{equation}
    \mathbf{d} = H_0\mathds{1}  + \boldsymbol{\varepsilon} + \mathbf{n}
\end{equation}
where $H_0 \sim \mathcal{N}\left(\mu_p, \sigma_p^2\right)$ incorporates the prior on $H_0$ and includes an uncertainty denoted by $\sigma_p^2$, $\boldsymbol{\varepsilon} \sim \mathcal{N}(\boldsymbol{\mu}_{\varepsilon, X_1 X_2 \dots}, \mathbf{C}_{\varepsilon, X_1 X_2 \dots})$ models a hypothetical bias, and $\mathbf{n} \sim \mathcal{N}(0, \mathbf{N})$ represents measurement noise, which takes account of the posterior uncertainty in the measurements of $H_0$. 


The Bayesian evidence is equal to the probability density of the data, evaluated at the measured values, when marginalised over all possible model parameter values. Under the assumptions listed above, we calculate this using basic properties of Gaussian random variables. To express this system in a compact way, we define

\begin{equation}
    \mathbf{x} =
\begin{pmatrix}
    H_0\mathds{1} \\
    \boldsymbol{\varepsilon} \\
    \mathbf{n}
\end{pmatrix}, \quad
    \mathbf{A} =
\begin{pmatrix}
    \mathbf{I} & \mathbf{I} & \mathbf{I}
\end{pmatrix}, \quad
    \mathbf{d} = \mathbf{A} \mathbf{x}
\end{equation}

\balance

The expected value and covariance of the data vector are then
\begin{equation}
    \langle \mathbf{d} \rangle = \mu_p \mathds{1} + \boldsymbol{\mu}_{\varepsilon, X_1 X_2 \dots}, \
    \mathbf{C_d} = \sigma_p^2 \mathds{1}\mathds{1}^T + \mathbf{C}_{\varepsilon, X_1 X_2 \dots} + \mathbf{N}
\end{equation}
Thus, the data are marginally distributed as
\begin{equation}
    \mathbf{d} \sim \mathcal{N}(\mu_p \mathds{1} + \boldsymbol{\mu}_{\varepsilon, X_1 X_2 \dots}, \mathbf{C_d})
\end{equation}
Based on our calculations above, the evidence for the model is given by
\begin{equation}
\begin{split}
    \mathcal{Z} = P(\mathbf{d}|\mathcal{M}_{X_1 X_2 \dots}) \
    = \frac{1}{\sqrt{|2\pi \mathbf{C_d}|}} \exp \left( -\frac{1}{2} (\mathbf{d} - \langle      \mathbf{d} \rangle)^T \mathbf{C_d}^{-1} (\mathbf{d} - \langle \mathbf{d} \rangle)       \right)
\end{split}
\end{equation}
Taking the logarithm gives the log evidence,
\begin{equation}
\begin{split}
    \ln \mathcal{Z} = -\frac{1}{2} \left[ \ln |2\pi \mathbf{C_d}| + (\mathbf{d} - \langle \mathbf{d} \rangle)^T \mathbf{C_d}^{-1} (\mathbf{d} - \langle \mathbf{d} \rangle) \right]
\end{split}
\end{equation}

This expression forms the basis for Bayesian model comparison under these priors. By computing and comparing the evidence for different models, each with different bias assumptions, we can identify which subset-specific models best describe the observed $H_0$ data.


\bsp	
\label{lastpage}
\end{document}